%% file: main.tex
\documentclass{article}
\usepackage{spconf}

\usepackage{amsmath, amssymb, amsfonts}
\usepackage{algorithmic}
\usepackage{graphicx}
\usepackage{makecell}
\usepackage{hyperref}
\usepackage{textcomp}
\usepackage{xcolor}
\usepackage{booktabs}
\usepackage{multirow}
\usepackage{array}

\usepackage{adjustbox}
\usepackage{kotex}
\usepackage{subcaption}
\usepackage{caption}
\usepackage[utf8]{inputenc}

\ninept
\begin{document}
\title{Forget who you Forgot: Speaker Unlearning to Prevent Re-Identification in Zero-Shot Text-to-Speech
}

\name{Hyoeun Kim, Yujun Lee, Kyuhong Shim}
\address{ Sungkyunkwan University, Republic of Korea \\
\texttt{\small \{hyoeunkim, yj090744, khshim\}@skku.edu} }

\maketitle

\begin{abstract}
    \input{section/00_abstract}
\end{abstract}


\input{section/01_introduction}
\input{section/02_related}
\input{section/03_method}
\input{section/04_experiment}
\input{section/05_conclusion}

\newpage

\section{Compliance with Ethical Standards}
This study used existing publicly available datasets, and no new data were collected. No ethical approval was required.

\bibliographystyle{IEEEbib}
\bibliography{reference}
\end{document}

%% file: section/00_abstract.tex
Recent zero-shot text-to-speech (ZS-TTS) systems can reproduce a speaker’s voice with high fidelity from only a few seconds of reference speech, raising concerns over unauthorized voice cloning and impersonation.
Speaker identity unlearning has recently emerged as an approach to selectively suppress this capability for speakers who opt out while preserving synthesis capability for other speakers.
Although existing approaches reduce speaker similarity, preventing re-identification often faces severe degradation of speech quality.
Motivated by this observation, we propose \textbf{GUARD}, a lightweight speaker identity unlearning framework that combines a learned speaker gate with speaker-agnostic activation steering on a frozen TTS backbone.
The steering vectors are optimized using group-relative reward optimization to shift outputs from forget speakers toward population-level impostor similarity while preserving intelligibility and speech naturalness.
On CosyVoice2, GUARD reduces forget-speaker similarity from 0.541 to 0.103 and re-identification accuracy in a 150-speaker gallery from 73.5\% to 0.5\%, while preserving retain-speaker reproduction.
The results demonstrate that similarity reduction alone may not fully characterize successful speaker identity unlearning and highlight re-identification as a complementary criterion for its evaluation.

%% file: section/01_introduction.tex
\section{Introduction}\label{sec:intro}

Zero-shot text-to-speech (ZS-TTS) systems reproduce the voices of previously unseen speakers from short reference recordings.
Models such as VALL-E~\cite{wang2023valle}, NaturalSpeech~\cite{shen2024naturalspeech2}, Voicebox~\cite{le2023voicebox}, F5-TTS~\cite{chen2024f5tts}, and CosyVoice~\cite{du2024cosyvoice2} synthesize natural speech that matches the input text while preserving speaker characteristics from the reference.
However, this capability raises the risks of unauthorized voice cloning, impersonation, voice phishing, and deepfake misuse.

Voice anonymization transforms speaker identity while preserving linguistic content~\cite{panariello2024voiceprivacy,tomashenko2026third}, whereas anti-cloning methods introduce adversarial perturbations into speech to disrupt unauthorized voice cloning~\cite{yu2023antifake,hu2026voicecloak}.
These approaches protect individual speech samples but do not address speaker opt-out requests in deployed TTS systems.
Speaker identity unlearning instead suppresses the reproduction of speakers who opt out after deployment.
Accordingly, we consider a provider-controlled deployment setting in which users access the TTS system through the standard synthesis interface.

Prior approaches follow two directions: TGU and SGU~\cite{kim2025not} fine-tune the flow decoder using alternative speech targets; TruS~\cite{lee2026erasing} removes speaker-identity-related directions from internal activations at inference time.
Recent work considers continual speaker identity unlearning under sequential opt-out requests~\cite{kim2026continual}.
Despite these advances, existing methods do not reduce speaker similarity (SIM) between generated and reference speech enough to suppress gallery-based re-identification.
We emphasize that this re-identification risk has received limited attention in prior speaker-unlearning work, and argue that speaker identity unlearning should suppress reliable re-identification while preserving linguistic content, speech naturalness, and retain-speaker reproduction.

\begin{figure}[t]
    \centering
    \includegraphics[width=\linewidth]{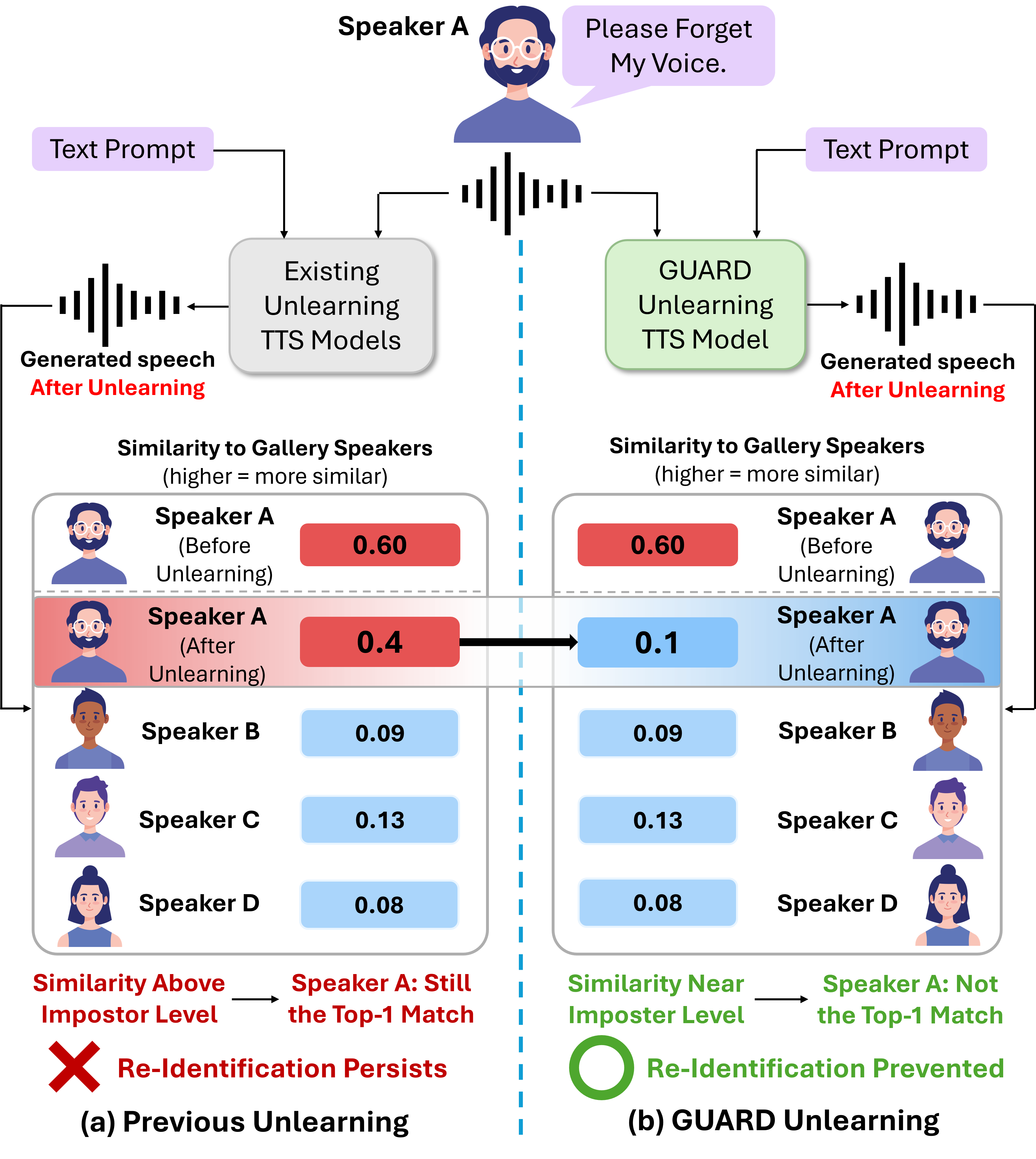}
    \caption{Motivation for re-identification-aware speaker unlearning: GUARD shifts forget-speaker similarity toward population-level impostor similarity.
    }
    \label{fig:motivation}
\end{figure}

\begin{figure*}[t]
    \centering
    \includegraphics[width=\linewidth]{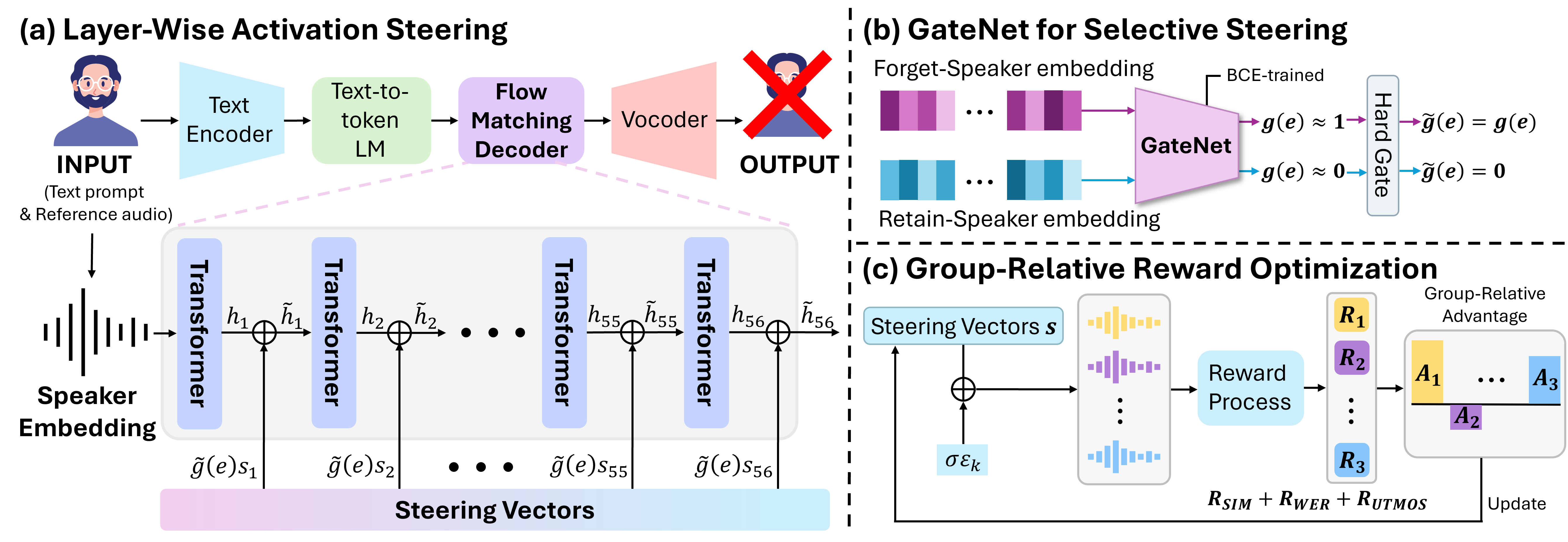}
    \vspace{0.1cm}
    \caption{GUARD architecture. (a) Layer-wise steering vectors modify the residual activations of the decoder transformer blocks. (b) GateNet determines whether steering is applied to the input speaker. (c) Group-relative reward optimization learns the steering vectors from post-generation rewards.}
    \label{fig:main}
\end{figure*}

To this end, we propose \textbf{GUARD} (Group-relative Unlearning with Activation Steering for Re-identification Defense), a re-identification-aware framework that targets the population-level impostor similarity.
We evaluate residual speaker identity through gallery-based re-identification.
GUARD combines GateNet with speaker-agnostic layer-wise steering vectors.
GateNet predicts the steering coefficient that weights the layer-wise steering vectors added to decoder activations.
We optimize these steering vectors using group-relative reward optimization with post-generation rewards that drive forget-speaker SIM toward population-level impostor similarity while preserving linguistic content and speech naturalness.
On CosyVoice2, GUARD reduces forget-speaker SIM from 0.541 to 0.103 and top-1 re-identification accuracy in a 150-speaker gallery from 73.5\% to 0.5\%, without degradation in speech quality and retain-speaker reproduction.

\vspace{0.1cm}
\noindent Our contributions are threefold:
\vspace{-0.1cm}
\begin{itemize}
    \setlength{\itemsep}{0pt}
    \setlength{\parskip}{0pt}
    \setlength{\parsep}{0pt}
    \setlength{\topsep}{3pt}
    \item We introduce a re-identification-aware formulation of speaker identity unlearning that targets population-level impostor similarity and evaluates residual identity through gallery-based re-identification.
    \item We propose GUARD, which combines a learned speaker gate with layer-wise activation steering and group-relative reward optimization on a frozen TTS backbone.
    \item GUARD substantially reduces gallery-based re-identification while preserving speech quality and retain-speaker reproduction, with lightweight roster updates and consistent improvements across other ZS-TTS backbones.
\end{itemize}

%% file: section/02_related.tex
\section{Related Work}
\label{sec:related}

\subsection{Speaker Identity Unlearning for Zero-Shot TTS}

Recent studies have investigated speaker identity unlearning for selectively suppressing specific speakers in zero-shot TTS.
TGU and SGU~\cite{kim2025not} optimize the flow decoder using alternative speech targets for forget speakers while preserving synthesis capability for retain speakers.
However, both methods require fine-tuning the full flow decoder and can affect retain-speaker reproduction.
TruS~\cite{lee2026erasing} instead suppresses speaker identity by steering internal activations along identity-related directions at inference time without additional training.
Increasing its steering strength reduces forget-speaker similarity but can substantially degrade speech naturalness and intelligibility, whereas weaker steering leaves substantial gallery-based re-identification.
Recent work also considers continual speaker identity unlearning under sequential opt-out requests~\cite{kim2026continual}.
Unlike prior work, we target population-level impostor similarity and explicitly evaluate residual speaker identity through gallery-based re-identification.

\subsection{Activation Steering for TTS}

Activation steering directly modifies internal representations of pretrained TTS models to control speech attributes.
EmoSteer-TTS~\cite{xie2025emosteer} derives activation-difference vectors for training-free emotion control.
EmoShift~\cite{zhou2026emoshift} learns trainable steering vectors for emotion-aware speech synthesis.
Activation steering has also been applied to accent modification while preserving speaker characteristics~\cite{yang2026activation}.
These studies demonstrate that internal activations provide a lightweight intervention space for controlling TTS generation.
GUARD instead optimizes speaker-agnostic layer-wise steering vectors for speaker identity unlearning using post-generation anonymity, intelligibility, and naturalness rewards.

\subsection{Reward-Based Optimization for TTS}

Reward-based optimization incorporates post-generation evaluation signals directly into TTS optimization.
Seed-TTS~\cite{anastassiou2024seed} uses reinforcement learning, and F5R-TTS~\cite{sun2025f5r} applies GRPO; both incorporate WER and speaker-similarity rewards.
GLASS~\cite{kang2026glass} optimizes lightweight LoRA-based acoustic style controls on a frozen TTS backbone using GRPO and post-generation acoustic metrics.
In contrast to these weight-space adaptations, GUARD performs group-relative reward optimization over layer-wise activation steering vectors while keeping the TTS backbone frozen.
Its anonymity reward drives forget-speaker similarity toward population-level impostor similarity, together with WER and UTMOS rewards for intelligibility and speech naturalness.

\input{tables/table_main_gallery}

%% file: tables/table_main_gallery.tex
\begin{table*}[t]
\centering

\setlength{\tabcolsep}{1pt}
\renewcommand{\arraystretch}{1.08}
\caption{
Quantitative results on LibriTTS clean-460 using CosyVoice2-0.5B. 
Gallery Re-ID reports top-1 re-identification accuracy over galleries of different sizes. 
For forget speakers, $\bar{\mu}_{\mathrm{imp}}=0.129$ denotes the mean population-level impostor similarity across forget speakers and serves as the reference for SIM; the optimal AUROC is 0.5, which indicates chance-level ranking for the evaluated detectability score.
}
\label{tab:main_results}
\resizebox{\linewidth}{!}{
\begin{tabular}{lcccccccccccc}
\toprule

\multirow{2}{*}{Method}
& \multicolumn{4}{c}{Forget Speakers (F)}
& \multicolumn{4}{c}{Retain Speakers (R)}
& \multicolumn{3}{c}{Gallery Re-ID (\%)$\downarrow$}
& \multicolumn{1}{c}{Imp. Det.} \\

\cmidrule(lr){2-5}
\cmidrule(lr){6-9}
\cmidrule(lr){10-12}
\cmidrule(lr){13-13}

& SIM $\rightarrow \bar{\mu}_{\mathrm{imp}}$
& WER $\downarrow$
& UTMOS $\uparrow$
& spk-ZRF $\uparrow$

& SIM $\uparrow$
& WER $\downarrow$
& UTMOS $\uparrow$
& spk-ZRF $\uparrow$

& id@50
& id@100
& id@150

& AUROC $\rightarrow$ 0.5
\\

\midrule

CosyVoice2
& 0.541 & 5.5 & 4.37 & 0.851
& 0.607 & 4.9 & 4.40 & 0.832
& 81.0 & 77.5 & 73.5
& 0.009 \\

SGU
& 0.481 & 5.6 & 4.40 & 0.856
& 0.556 & 5.2 & 4.41 & 0.838
& 67.5 & 63.0 & 56.5
& 0.021 \\

TGU
& 0.481 & 5.7 & 4.38 & 0.861
& 0.561 & 5.7 & 4.39 & 0.840
& 64.5 & 62.5 & 56.0
& 0.021 \\

TruS ($\alpha=2.0$)
& 0.341 & 5.3 & 3.64 & 0.837
& 0.602 & 4.6 & 4.39 & 0.832
& 34.5 & 30.0 & 25.0
& 0.117 \\

TruS ($\alpha=2.5$)
& 0.078 & 122.0 & 1.35 & 0.807
& 0.614 & 4.1 & 4.38 & 0.832
& 3.5 & 2.5 & 2.0 
& 0.657 \\

\textbf{GUARD (Ours)}
& 0.103 & 5.9 & 4.37 & 0.824
& 0.601 & 4.7 & 4.40 & 0.831
& \textbf{2.0} & \textbf{1.0} & \textbf{0.5}
& \textbf{0.579} \\

\bottomrule
\end{tabular}
}
\end{table*}

%% file: section/03_method.tex
\section{Method}
\label{sec:method}

\subsection{Problem Formulation and Overview}
\label{sec:formulation}

A pretrained zero-shot TTS model $\mathcal{M}$ synthesizes speech from a reference recording $a$ and target text $y$.
Given a set of forget speakers $\mathcal{F}$, GUARD suppresses their re-identification while preserving linguistic content, speech naturalness, and retain-speaker reproduction.
We use CosyVoice2~\cite{du2024cosyvoice2} as the primary backbone and freeze all backbone parameters.
Rather than minimizing speaker similarity, GUARD drives it toward population-level impostor similarity, which characterizes the similarity between unrelated speakers.

GUARD modifies residual activations in the flow decoder using layer-wise steering vectors.
For decoder block $\ell$, we apply
\begin{equation}
\tilde{\mathbf{h}}_{\ell}
=
\mathbf{h}_{\ell}
+
\tilde{g}(\mathbf{e})\mathbf{s}_{\ell},
\qquad
\ell=1,\ldots,L,
\label{eq:steering}
\end{equation}
at each CFM step and temporal position.
$\mathbf{s}_{\ell}\in\mathbb{R}^{d}$ is a trainable steering vector shared across forget speakers, and $\tilde{g}(\mathbf{e})$ controls its contribution for the input speaker.
CosyVoice2 has $L=56$ flow-decoder blocks with hidden dimension $d=256$, for a total of 14.3K trainable steering parameters.

\subsection{GateNet for Selective Steering}
\label{sec:gate}

GateNet predicts the steering coefficient from the speaker embedding using a lightweight MLP with SiLU activation.
The network uses a 128-dimensional hidden layer and contains 24.8K parameters.
We train GateNet with class-weighted binary cross-entropy using forget and retain speaker labels.
At inference time, we apply an asymmetric hard threshold:
\begin{equation}
\tilde{g}(\mathbf{e})
=
\begin{cases}
0, & g(\mathbf{e})<\tau,\\
g(\mathbf{e}), & \mathrm{otherwise},
\end{cases}
\qquad \tau=0.3.
\label{eq:hard_gate}
\end{equation}
When $g(\mathbf{e})<\tau$, the hard gate removes steering entirely, preserving the original computation path of the frozen backbone.
GateNet is trained separately and remains fixed during steering-vector optimization.

\subsection{Group-Relative Reward Optimization}
\label{sec:optimization}

We optimize the layer-wise steering vectors using post-generation rewards for anonymity, intelligibility, and speech naturalness.
Let $\mathbf{s}$ denote the complete set of layer-wise steering vectors.
For each input, we generate $G$ candidate sets of steering vectors as
$\mathbf{s}^{(k)}=\mathbf{s}+\sigma\boldsymbol{\epsilon}_k$,
where $\boldsymbol{\epsilon}_k\sim\mathcal{N}(\mathbf{0},\mathbf{I})$.
Each candidate produces a synthesized output $x^{(k)}$, which is scored using the following reward:
\begin{equation}
\begin{aligned}
R(x^{(k)}) ={}&
-w_a
\left|
\cos\!\left(E_{\mathrm{sv}}(x^{(k)}),\mathbf{p}_{f}\right)
-\mu_{\mathrm{imp}}(f)
\right|
\\[-1mm]
&-w_c\,\mathrm{WER}(x^{(k)},y)
+w_n\min\left(0,U(x^{(k)})-U_0\right),
\end{aligned}
\label{eq:reward}
\end{equation}
where $E_{\mathrm{sv}}$ denotes the speaker-verification encoder,
$\mathbf{p}_{f}$ denotes the enrollment prototype of forget speaker $f$,
and $\mu_{\mathrm{imp}}(f)$ denotes its population-level impostor similarity,
computed as the mean cosine similarity between $\mathbf{p}_{f}$ and non-matching speaker prototypes.
The anonymity term penalizes deviation of forget-speaker SIM from $\mu_{\mathrm{imp}}(f)$.
The intelligibility term penalizes transcription errors using WER, while the speech naturalness term penalizes outputs whose UTMOS score falls below $U_0$, the baseline's UTMOS score.

We normalize rewards within each candidate group and estimate the reward gradient as
\begin{equation}
A_k
=
\frac{R_k-\bar{R}}{\mathrm{std}(R)+\delta},
\qquad
\nabla_{\mathbf{s}}\widehat{J}
=
\frac{1}{G\sigma}
\sum_{k=1}^{G}
A_k\boldsymbol{\epsilon}_k.
\label{eq:gradient}
\end{equation}
We update the steering vectors with Adam~\cite{kingma2014adam} using forget-speaker candidates only, while GateNet remains fixed.
At inference, GUARD applies the learned steering vectors with the gated coefficient $\tilde{g}(\mathbf{e})$, without candidate sampling or reward computation.

%% file: section/04_experiment.tex
\section{Experiments}
\label{sec:experiments}

\subsection{Experimental Settings}
\label{sec:settings}

\vspace{0.1cm}\noindent\textbf{Dataset and implementation.}
We evaluate GUARD on LibriTTS clean-460~\cite{zen2019libritts}, designating 50 speakers (25 male and 25 female) as forget speakers and the remaining 995 as retain speakers.
Evaluation uses 200 utterances from the 50 forget speakers and 80 utterances from 40 retain speakers.
All evaluation utterances are held out from GateNet training, and the 200 forget-speaker evaluation utterances are also held out from steering-vector optimization.
Each synthesis example uses a 3-second speaker prompt from one utterance and target text from a different utterance of the same speaker.
We use the frozen CosyVoice2-0.5B~\cite{du2024cosyvoice2} as the primary backbone.
For backbone generalization, we additionally evaluate F5-TTS~\cite{chen2024f5tts} and FireRedTTS~\cite{guo2024fireredtts}.
For both models, we use an external speaker-verification encoder to obtain GateNet inputs and optimize the steering vectors separately for each backbone.

GateNet is trained separately on cached speaker embeddings for 2.5K steps.
We optimize the layer-wise steering vectors over the 50 forget speakers while keeping GateNet fixed.
Steering-vector optimization uses $G=16$ candidates with $\sigma=0.06$, reward weights $(w_a,w_c,w_n)=(1,24,4)$, and $U_0=4.45$.
We use Whisper~\cite{radford2023robust} for the WER reward computation and performance evaluation.
All experiments are conducted on a single V100-16GB GPU.

\vspace{0.1cm}\noindent\textbf{Baselines and metrics.}
We compare GUARD with the original CosyVoice2 and three speaker identity unlearning methods: SGU, TGU, and TruS.
SGU and TGU fine-tune the full 71.3M-parameter flow decoder using their original objectives with $\lambda=0.2$, 60 utterances per forget speaker, and 12K optimization steps.
We evaluate TruS at steering strengths $\alpha=2.0$ and $2.5$ with oracle gating for retain speakers.

We report speaker similarity (SIM)~\cite{chen2022wavlm}, WER~\cite{radford2023robust}, UTMOS~\cite{saeki2022utmos}, and spk-ZRF~\cite{kim2025not} for both forget and retain speakers.
We additionally measure residual speaker identity using gallery-based top-1 re-identification accuracy, $\mathrm{id}@N$, for $N\in\{50,100,150\}$.
For each synthesized forget-speaker utterance, the gallery contains the corresponding forget speaker and $N-1$ distractors, each represented by an enrollment prototype.
We also report detectability AUROC between protected outputs and genuine impostor recordings, where values closer to $0.5$ indicate lower detectability.
Averaging $\mu_{\mathrm{imp}}(f)$ over the 50 forget speakers gives $\bar{\mu}_{\mathrm{imp}}=0.129$.

\subsection{Results}
\label{sec:results}

\input{tables/table_ablation}
\input{tables/table_cost}
\input{tables/table_others}

\noindent\textbf{Speaker unlearning and re-identification.}
Table~\ref{tab:main_results} shows that GUARD shifts forget-speaker SIM toward the population-level impostor regime while preserving retain-speaker reproduction.
SGU and TGU update the full flow decoder; their re-identification rates at $\mathrm{id}@150$ remain at $56.5\%$ and $56.0\%$, respectively.
TruS reduces forget-speaker SIM as the steering strength $\alpha$ increases; at $\alpha=2.0$, its re-identification rate at $\mathrm{id}@150$ remains at $25.0\%$.
At $\alpha=2.5$, the re-identification rate of TruS decreases to $2.0\%$, but WER increases to $122.0\%$ and UTMOS decreases to $1.35$, which indicates a severe performance degradation.

GUARD reduces forget-speaker SIM from $0.541$ to $0.103$ and $\mathrm{id}@150$ from $73.5\%$ to $0.5\%$, while maintaining WER, UTMOS, and retain-speaker metrics close to the original backbone.
Its detectability AUROC of $0.579$ is closest to $0.5$ among the evaluated methods, indicating that protected outputs are less distinguishable from genuine impostor recordings.

\vspace{0.15cm}\noindent\textbf{Reward ablation.}
Table~\ref{tab:ablation} examines the effect of each reward term on forget-speaker identity suppression, intelligibility, and speech naturalness.
Removing the impostor target (i.e., $\mu_{\text{imp}}= 0$) lowers forget-speaker SIM from $0.103$ to $0.056$, but does not reduce gallery-based re-identification and increases detectability AUROC from $0.579$ to $0.706$.
This result indicates that reducing forget-speaker SIM below the population-level impostor regime does not further reduce gallery-based re-identification and makes protected outputs more distinguishable from genuine impostor recordings.
Removing the UTMOS reward substantially degrades both speech naturalness and intelligibility, with UTMOS decreasing to $1.25$ and WER increasing to $11.4\%$.
Without the WER reward, WER increases to $6.7\%$, while $\mathrm{id}@150$ increases to $2.5\%$.

\vspace{0.15cm}\noindent\textbf{Update cost and backbone generalization.}
Table~\ref{tab:enroll_cost} compares the cost of adding a newly opted-out speaker to the forget roster.
GUARD keeps the speaker-agnostic layer-wise steering vectors fixed as the forget roster changes and updates only the 24.8K-parameter GateNet, without requiring TTS-backbone retraining.
This lightweight roster update requires 9.9 minutes, compared with 6.3--7.8 hours for SGU and TGU, which fine-tune the 71.3M-parameter flow decoder.
The lightweight roster update also requires substantially less GPU memory and checkpoint storage.

Table~\ref{tab:model_generalization} evaluates GUARD on F5-TTS~\cite{chen2024f5tts} and FireRedTTS~\cite{guo2024fireredtts}.
GUARD reduces $\mathrm{id}@150$ from $95.5\%$ to $6.0\%$ on F5-TTS and from $91.5\%$ to $18.0\%$ on FireRedTTS while maintaining intelligibility and speech naturalness.

\subsection{Analysis}
\label{sec:analysis}

\noindent\textbf{Population-level impostor similarity as the target.}
Reducing forget-speaker SIM alone does not fully characterize speaker identity unlearning.
SIM above the population-level impostor regime can retain speaker-specific information that supports gallery-based re-identification.
Reducing SIM below this regime does not further reduce gallery-based re-identification and makes protected outputs more distinguishable from genuine impostor recordings.
These observations motivate population-level impostor similarity as a reference target rather than unrestricted SIM minimization.

\vspace{0.15cm}\noindent\textbf{Selective intervention and roster updates.}
GUARD separates speaker selection from the identity intervention.
GateNet determines whether steering is applied, while the speaker-agnostic layer-wise steering vectors define the shared activation intervention for forget speakers.
This separation also localizes roster updates to GateNet, so newly opted-out speakers can be incorporated without requiring TTS-backbone retraining.

%% file: tables/table_ablation.tex
\begin{table}[t]
\centering
\setlength{\tabcolsep}{0.8pt}
\renewcommand{\arraystretch}{1.0}
\caption{
Reward ablation of GUARD with all variants trained from scratch under an identical budget.
Each row removes one reward term to isolate its effect on unlearning performance.
The optimal AUROC score is 0.5.
}
\label{tab:ablation}
\vspace{-0.1cm}
\resizebox{1.0\linewidth}{!}{
\begin{tabular}{lccccccc}
\toprule
\multirow{2}{*}{Variant}
& \multicolumn{3}{c}{Forget Speakers (F)}
& \multicolumn{3}{c}{Gallery Re-ID (\%) $\downarrow$}
& \multicolumn{1}{c}{Imp. Det.} \\
\cmidrule(lr){2-4}
\cmidrule(lr){5-7}
\cmidrule(lr){8-8}
& SIM $\rightarrow \bar{\mu}_{\mathrm{imp}}$
& WER $\downarrow$
& UTMOS $\uparrow$
& id@50
& id@100
& id@150
& AUROC \\
\midrule
\textbf{GUARD}
& 0.103 & 5.9 & 4.37 & 2.0 & 1.0 & 0.5 & 0.579 \\
\quad w/o $\mu_{\mathrm{imp}}$
& 0.056 & 6.2 & 4.39 & 4.5 & 2.0 & 2.0 & 0.706 \\
\quad w/o $R_{\text{UTMOS}}$
& 0.154 & 11.4 & 1.25 & 3.5 & 2.5 & 1.0 & 0.430 \\
\quad w/o $R_{\text{WER}}$
& 0.092 & 6.7 & 4.44 & 6.5 & 3.0 & 2.5 & 0.608 \\
\bottomrule
\end{tabular}
\vspace{0.2cm}
}
\end{table}

%% file: tables/table_cost.tex
\begin{table}[t]
\centering
\setlength{\tabcolsep}{7.0pt}
\renewcommand{\arraystretch}{1.0}
\caption{Cost of updating a single newly opted-out speaker.}
\label{tab:enroll_cost}
\vspace{-0.1cm}
\resizebox{1.0\linewidth}{!}{
\begin{tabular}{lcccc}
\toprule
Method & Updated params & Update time & Memory & Update size \\
\midrule
TGU   & 71.3M & 6.3\,h & 8--12\,GB & 286\,MB \\
SGU   & 71.3M & 7.8\,h & 8--12\,GB & 286\,MB \\
\textbf{GUARD} & \textbf{24.8K} & \textbf{9.9\,min} & \textbf{$<$1\,GB} & \textbf{0.1\,MB} \\
\bottomrule
\end{tabular}
}
\end{table}

%% file: tables/table_others.tex
\begin{table}[t]
\centering
\caption{
Backbone generalization of GUARD across F5-TTS and FireRedTTS.
For forget-speaker SIM, $\bar{\mu}_{\mathrm{imp}}=0.129$ denotes the mean population-level impostor similarity across forget speakers.
}
\vspace{-0.1cm}
\label{tab:model_generalization}
\setlength{\tabcolsep}{1.0pt}
\renewcommand{\arraystretch}{1.05}
\resizebox{\linewidth}{!}{
\begin{tabular}{llcccccc}
\toprule
\multirow{2}{*}{Backbone}
& \multirow{2}{*}{Method}
& \multicolumn{3}{c}{Forget Speakers (F)}
& \multicolumn{3}{c}{Gallery Re-ID (\%) $\downarrow$} \\
\cmidrule(lr){3-5}
\cmidrule(lr){6-8}
&
& SIM $\rightarrow \bar{\mu}_{\mathrm{imp}}$
& WER $\downarrow$
& UTMOS $\uparrow$
& id@50
& id@100
& id@150 \\
\midrule

\multirow{2}{*}{F5-TTS}
& Baseline
& 0.651 & 3.5 & 4.13
& 96.5 & 95.5 & 95.5 \\
& GUARD
& \textbf{0.199} & 4.2 & 4.25
& \textbf{14.5} & \textbf{6.5} & \textbf{6.0} \\

\midrule

\multirow{2}{*}{FireRedTTS}
& Baseline
& 0.636 & 4.2 & 4.17
& 94.5 & 93.0 & 91.5 \\
& GUARD
& \textbf{0.210} & 4.8 & 4.15
& \textbf{26.0} & \textbf{20.0} & \textbf{18.0} \\

\bottomrule
\end{tabular}
\vspace{0.2cm}
}
\end{table}

%% file: section/05_conclusion.tex
\section{Conclusion}
\label{sec:conclusion}

We introduced GUARD, a speaker identity unlearning framework that combines GateNet with speaker-agnostic layer-wise steering vectors optimized using group-relative reward optimization on a frozen ZS-TTS backbone.
By driving forget-speaker SIM toward the population-level impostor regime, GUARD substantially reduced gallery-based re-identification while preserving speech quality and retain-speaker reproduction.
The speaker-agnostic layer-wise steering vectors remained fixed during roster updates, requiring only GateNet retraining for newly opted-out speakers.
Future work will investigate speaker-adaptive steering vectors for individualized speaker identity suppression while preserving GUARD's lightweight update mechanism.